\documentclass[aps,prl,showpacs,twocolumn]{revtex4-1}
\usepackage{epsfig}
\usepackage{color}

\definecolor{red}{rgb}{1,0,0}
\definecolor{blue}{rgb}{0,0,1}

\newcommand{\be}{\begin{equation}}
\newcommand{\ee}{\end{equation}}
\newcommand{\ba}{\begin{eqnarray}}
\newcommand{\ea}{\end{eqnarray}}

\begin{document}
\title{Stability at  Random Close Packing}

\author{ Matthieu Wyart }

\affiliation{New York University, Center for Soft Matter Research, 4 Washington Place, New York, NY, 10003, USA }
\date{\today}

\begin{abstract}

The requirement that  packings of hard particles,  arguably the simplest structural glass, cannot be compressed by rearranging their network of contacts   is shown to yield a new constraint on their microscopic structure. This constraint takes the form  a bound between the 
distribution of contact forces $P(f)$ and the pair distribution function $g(r)$:  if $P(f)\sim f^\theta$ and $g(r)\sim(r-\sigma_0)^{-\gamma}$, where $\sigma_0$ is the particle diameter,  one finds that $\gamma\geq1/(2+\theta)$.  
This bound plays a role similar to those found in some glassy  materials with long-range interactions, such as the Coulomb gap in Anderson insulators or the distribution of local fields in  mean-field spin glasses. There is ground to believe that this bound is saturated, offering an explanation for the presence of avalanches of rearrangements with power-law statistics observed in packings.

\end{abstract}


\maketitle


Amorphous materials are perhaps the simplest example of glasses, in which the dynamics is so slow that thermal equilibrium cannot be reached.  In these systems properties are history-dependent, and configurations of equal energy are not equiprobable. What principles  then govern which part of the configuration space is explored, for example when a pile of sand is prepared? One approach was proposed by Edwards in the context of granular matter \cite{edwards}, and is based on the hypothesis that all mechanically stable states are equiprobable. Another line of thought assumes that the configurations generated by the dynamics are linearly stable, but only marginally \cite{Wyart052,Wyart053}: the microscopic structure is such that soft elastic modes are present at vanishingly small frequencies. This view can explain  \cite{Wyart052,Wyart053,brito3} in particular the singularities occurring in the coordination number and in the elasticity of amorphous solids made of repulsive particles near the unjamming threshold \cite{O'hern03, Silbert05,revue} where rigidity disappears. Despite these successes, the hypothesis of linear marginal stability yields an incomplete insight on the non-linear processes occurring in amorphous materials, which are critical to understand plasticity, thermal activation or  granular flows \cite{revue}. When interactions are short-range one key source of non-linearity is the creation or destruction of contacts between particles \cite{rouxDL,Schreck}. Combe and Roux have observed numerically \cite{rouxDL} that such rearrangements occur intermittently, in bursts or avalanches whose size  is power-law distributed, a kind of dynamics referred to as crackling noise \cite{crack}.  

Interestingly some glassy systems with long-range interactions display such dynamics, in particular Coulomb glasses \cite{monroe} and mean-field spin glasses \cite{Pazmandi}. In both cases the requirement of stability toward  discrete excitations (flipping two spins or moving one electron) leads to bounds on important physical quantities:  Efros and Shklovskii  showed that the density of states in a Coulomb glass must vanish at the Fermi energy \cite{efros}, implying the presence of the so-called Coulomb gap.  Thouless, Anderson and Palmer \cite{thouless} demonstrated for mean-field spin glasses that the distribution of local fields must vanish at least linearly at low fields. In these systems  the near saturation of the stability bound strongly affect physical properties, and is responsible for the crackling noise.

In this letter I argue that the same scenario holds in packings of hard frictionless spheres. I derive a stability bound toward discrete excitations, associated with the opening and the closing of contacts. This bound constrains
the pair distribution function $g(r)$ and the distribution $P(f)$ of the magnitude of contact forces $f$ between particles. The presence of weak forces is found to destabilize the system, whereas the abundance of pairs of particles that are very close to each other but not touching stabilizes it. If $\sigma_0$ is the particle diameter and $g(r)$ and $P(f)$ are assumed to obey power laws, $g(r)\sim (r-\sigma_0)^{-\gamma}$ and $P(f)\sim f^\theta$, I find that stability implies $\gamma\geq1/(2+\theta)$. 

There is ground to believe that the contact networks of packings are marginally stable, as previous observations, although incomplete, are consistent with the saturation of this bound. These results build a new link between  structural glasses and glasses with frozen disorder  \cite{kurchan,Zamponi}  where theoretical progress on avalanches has recently been made \cite{LeDoussal}, thus providing a new conceptual handle to investigate the rewiring of the contact network, of key importance for flow and plasticity.   Finally, these results  enable to relate the density landscape in sphere packings  to their microscopic structures, and suggest that some aspects of the much studied distribution of force  \cite{Liu2} cannot be captured by simple local arguments, but are rather controlled by subtle correlations in the structure associated with network stability.

I consider a packing of $N$ hard frictionless particles of diameter $\sigma_0$, in spatial dimension $d$. The packing is contained in a cubic box of  volume $V$ made of rigid walls, and is formed  by pushing particles together by reducing the box size, so as to apply a pressure $p$.  Microscopically, the boundaries apply external forces ${\vec F}_i$  on all the particles $i$ in contact with it.  
Mechanical stability requires that no floppy modes exist, apart from global translations and rotations.
Floppy modes are collective motions of the degrees of freedom of the system (that include the $Nd$  degrees of freedom of the particles and  changes in the box size) 
for which the distances between objects in contact (including both particles and the box) are fixed. If such a mode existed, the system would flow along it. 
Packings of hard spherical particles are {\it isostatic}:  the average number of contacts between particles, the coordination number $z$, is 
just sufficient to guarantee mechanical stability and to avoid the presence of floppy modes, corresponding to $z=z_c=2d$. 
It can be shown that this condition is necessary to ensure no overlaps between particles \cite{alexander,tkachenko2,moukarzel}.

In an isostatic system the removal of any contact leads to the creation of one floppy mode. Floppy modes can be generated as follows:   two particles $1$ and $2$, forming a contact labelled $\langle 12\rangle$, are pushed apart while all the other contacts remain closed. I denote by $\delta {\vec R}^{\langle 12\rangle}_{i}(s)$ the displacement of particle $i$ following the opening of the contact $\langle 12\rangle$ by a distance $s$. This displacement field is uniquely defined, because only one floppy mode appears when a contact is broken, and exists for $s$ sufficiently small, so as to ensure that no new contacts are formed in the system. Below I shall make one  hypothesis and use the following properties of floppy modes,  valid at random close packing:
(a) in packings of particles,  floppy modes extend in general in the entire system, and displace an extensive number of particles. If 
\be
\label{-1}
B_{kl}=\hbox{lim}_{s\rightarrow 0} \sum_i [\delta {\vec R}^{\langle kl\rangle}_{i}(s)]^2/(N s^2),
\ee
then $B= $ lim$_{N\rightarrow \infty} \langle B_{kl} \rangle>0$, where the average is made on all contacts. Intuitively this property stems from the fact that mechanical stability is not a local property,
but is governed by the mean coordination $z$, as shown by Maxwell \cite{maxwell}.  The floppy mode must explore the entire system to ``feel" that the average coordination is precisely $z_c$. To see this, consider an isostatic elastic network of spring of stiffness $k$ with one extra contact. These two aspects (finite $k$ and the additional spring)  confer a finite elasticity to the system, allowing to track where the elastic energy is propagating, and are not expected to change the statistical features of the displacement in response to a local strain. The energy stored in a  contact $ij$ after $12$ is stretched can be shown to be \cite{Wyart053}:
\be
\label{-1}
\delta E_{ij}=\frac{1}{2} k s^2 f_{ij}^2 f_{12}^2,
\ee
where $f_{ij}$ is the  force in the contact $ij$, normalized such that $\sum_{ij} f_{ij}^2=1$. In a sphere packing at some pressure mechanical stability implies that contact forces pervade the system, as observed, implying that the elastic energy and therefore the floppy mode is extended.  (b) The argument below focuses on floppy mode associated with contacts carrying a weak force. Although  lim$_{N\rightarrow \infty} \langle B_{kl} \rangle>0$ for a typical contact, it might not hold for the weakest contacts, and we may assume more generally that $B(f)\equiv \langle B_{kl}\rangle_{f}\sim f^\delta$, where the average is on all contact $\langle kl\rangle$ whose force is $f$.  I shall present the argument for the simplest assumption $\delta=0$, the extension to finite $\delta$ is straightforward and reported below. (c) For a floppy mode, the relative displacement between two adjacent particles is of order of the displacement of either particle. If two particles are moved apart in a normal (non-isostatic) elastic medium, this property would be true only for particles close to the chosen pair, and violated in the far field where the strain becomes much smaller than the displacement. In an isostatic system however Eq.(\ref{-1}) implies that the repartition of the energy, and therefore the properties of the displacement field, are independent from the distance to the source. Property (c) was  confirmed numerically for the lowest modes of isostatic packings \cite{Wyart052}. (d)  The response to a  force dipole of amplitude $F$ applied on two non-contacting particles extends to the entire system. In particular, the resulting change of amplitude of contact forces between particles is of order $F$ everywhere. This property can be derived formally from properties (a) and (c)  \cite{footnote1},  using the existence of a duality between floppy modes and force propagation  \cite{tkachenko1,tkachenko2}, and is supported by numerics \cite{respprl}.

A packing of hard particles has an infinite energy if particles overlap, and no energy otherwise. We shall focus on non-overlapping configurations, where the relevant energy is simply  $pV$. 
I shall argue that the stability of a packing against compression leads to constraints on the  packing geometry. 
Consider the floppy mode $\delta {\vec R}^{\langle 12\rangle}_{i}(s)$.
 The constraint that the change of distance $\delta r_{\langle ij \rangle}$ between particles in contact is null, except for the contact $\langle 12\rangle$,  can be expressed  at the second order using Pythagoras theorem as:
 \ba
 \label{0}
&\forall& \ {\langle ij \rangle}\neq\langle 12\rangle,\  \delta r_{\langle ij \rangle}=(\delta {\vec R}^{\langle 12\rangle}_{j}(s)-\delta{\vec R}^{\langle 12\rangle}_{i}(s))\cdot{\vec n}_{\langle ij \rangle}\nonumber \\
&+& \frac{[(\delta {\vec R}^{\langle 12\rangle}_{j}(s)-\delta{\vec R}^{\langle 12\rangle}_{i}(s))\cdot{\vec n}_{\langle ij \rangle}^\bot]^2}{2\sigma_0}+o(s^2)=0
 \ea
 where  ${\vec n}_{\langle ij \rangle}$ is the unit vector going from $i$ to $j$ in the initial configuration, and the notation $\cdot {\vec n}_{\langle ij \rangle}^\bot$ indicate the projection onto the space orthogonal to  ${\vec n}_{\langle ij \rangle}$.
 
 We now compute the change of volume associated with the displacement field $\delta {\vec R}^{\langle 12\rangle}_{i}(s)$. 
 Force balance in the unperturbed state can be written:
 \be
 \label{1}
 \forall i, \ \  {\vec F}_i-\sum_{ j(i)} f_{\langle ij \rangle} {\vec n}_{\langle ij \rangle}=0
\ee
where the sum is on all particles $j(i)$ in contact with $i$, ${\vec F}_i$ is the force exerted by the wall on particle $i$ (and is thus zero for most of the particles), and $f_{\langle ij \rangle}>0$ is the  magnitude of the force in the contact $\langle ij \rangle$. Multiplying Eq.(\ref{1}) by any displacement field $\delta {\vec R}_i$ and summing on all particles leads to the virtual work theorem:
\be
\label{2}
\sum_i {\vec F}_i\cdot \delta {\vec R}_i+\sum_{\langle ij \rangle} (\delta {\vec R}_j-\delta{\vec R}_i)\cdot{\vec n}_{\langle ij \rangle} f_{\langle ij \rangle}=0,
\ee
where the second sum is made on all contacts. In our system external forces only stem from the boundaries, and the associated work is $\sum_i {\vec F}_i\cdot\delta {\vec R}_i=-p\delta V$. Using this result, together with Eq.(\ref{0}) and Eq.(\ref{2}) applied to the floppy mode $\delta {\vec R}^{\langle 12\rangle}_i(s)$, one obtains:
\be
\label{3}
p\delta V(s)=s f_{\langle 12\rangle}- C +o(s^2)
\ee
where 
\be
\label{4}
C=\sum_{\langle ij \rangle\neq\langle 12\rangle} f_{\langle ij \rangle} \frac{[(\delta {\vec R}^{\langle 12\rangle}_{j}(s)-\delta{\vec R}^{\langle 12\rangle}_{i}(s))\cdot{\vec n}_{\langle ij \rangle}^\bot]^2}{\sigma_0}.
\ee
According to the properties  (a,c), $[(\delta {\vec R}^{\langle 12\rangle}_{j}(s)-\delta{\vec R}^{\langle 12\rangle}_{i}(s))\cdot{\vec n}_{\langle ij \rangle}^\bot]^2\sim s^2$. Thus $C= s^2 A N \langle f\rangle/\sigma_0$, where $\langle f\rangle$ is the average contact force and $A$ is a constant of order one. Eq.(\ref{3}) becomes:
\be
\label{3bis}
p\delta V(s)=s f_{\langle 12\rangle}- \frac{A N \langle f\rangle s^2}{\sigma_0}+o(s^2).
\ee

\begin{figure}[htb]
\begin{center} { 
\centerline{ \includegraphics[width=0.50\textwidth]{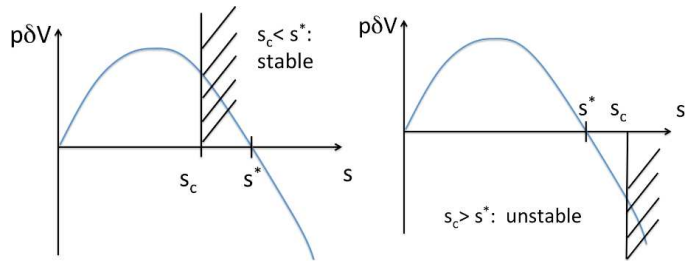} }
\caption{  \label{f1}  Energy change $p\delta V$ after opening a contact $\langle 12\rangle$ by a distance $s$. $s_c$ indicates the distance at which a new contact is formed, and motion along the considered floppy mode becomes impossible. If $s_c<s^*$ (Left) no denser state can be obtained by opening the contact $\langle 12\rangle$ (note, however, that a looser  but still metastable configuration can be obtained if $s_c$ is larger than the distance $s_{max}$ at which $p\delta V$ is a maximum).  If $s_c>s^*$ (Right), a denser, more stable state can be generated.
}}
\end{center}
\end{figure}

Eq.(\ref{3bis}) is plotted in Fig.(\ref{f1}).  Since the inter-particle potential is purely repulsive, $f_{\langle 12\rangle}>0$ for all $\langle 12\rangle$, implying that for sufficiently small $s$ opening a contact always increases $V$.   However the quadratic  term is always destabilizing.  A  denser state will thus be generated if the contact $\langle 12\rangle$ can be opened up to a distance $s>s^*$ without a new contact being formed, with:
\be
\label{3ter}
 s^*\sim \frac{f_{\langle 12\rangle}} {\langle f\rangle}\frac{\sigma_0}{ N},
 \ee
as indicated in the right panel of Fig.(\ref{f1}).  The initial configuration is stable however if a new contact is formed at some $s_c<s^*$, as motion along the floppy mode beyond $s_c$ is then forbidden. 

According to Eq.(\ref{3bis}), the most stringent constraint on stability corresponds to the opening of contacts with the weakest forces. I assume that the distribution of contact force $P(f)$ follows $P(f)\sim f^\theta/\langle f\rangle^{\theta+1}$ at low forces, where the term $\langle f\rangle^{\theta+1}$ ensures proper normalization.  I define the typical  smallest contact force $f_{min}$ in a system with  $N_c\equiv zN/2=Nd$ contacts  as the force magnitude for which there is in average one smaller contact force in the system:
\be
\label{5}
\int_0^{f_{min}} P(f) df \equiv 1/N_c\sim 1/N
\ee
leading to  $f_{min}\sim \langle f\rangle N^{-1/(1+\theta)}$. This estimation assumes that forces can be treated as independent variables, which I expect to be approximatively true. Using this force in Eq.(\ref{3ter}), which applies to contact with low-forces according to (b), one finds that  it is sufficient to open the contact with the weakest contact force by an amount  $s^*_{min}$ that satisfies:
\be
\label{4bis}
s^*_{min}/\sigma_0\sim\frac{ f_{min} }{ N \langle f\rangle}\sim  N^{-(2+\theta)/(1+\theta)},
\ee
to generate a denser packing. 

In a stable packing a new contact must be formed for some $s_c<s^*_{min}$. We now estimate the value of $s_c$ in terms of the pair distribution function $g(r)$. The first contact to form will correspond to particles that were almost touching in the initial configuration ($s=0$). We denote by $h_{min}$ the typical smallest separation between particles that are not in contact. According to properties (a,b,c),  the relative motion of nearby particles in a floppy mode is of order $s$. Thus the first contact will be formed for $s_c\sim h_{min}$, which can be expressed in terms of the pair distribution function:
\be
\label{6}
\int_{\sigma_0}^{\sigma_0+h_{min}}g(r)dr\equiv 1/N_c\sim 1/N
\ee
Assuming that $g(r)\sim (r-\sigma_0)^{-\gamma}$, one finds that $s_c/\sigma_0\sim h_{min}/\sigma_0\sim N^{-1/(1-\gamma)}$.

 The stability conditions  $s_c< s^*_{min}$ thus implies that $N^{-1/(1-\gamma)}< N^{-(2+\theta)/(1+\theta)}$, or equivalently:
 \be
 \label{7}
 \gamma\geq\frac{1}{2+\theta}
 \ee
which is my main result. For $\delta\neq0$ the same argument leads to $\gamma\geq (1-\delta/2)/(2+\theta-\delta/2)$.

I now show that if the inequality (\ref{7}) is violated, opening the contact with one of the smallest contact forces would lead to a giant avalanche that restructures an extensive number of contacts in the system. 
Let us denote by ${\langle 34\rangle}$ the contact that closes at $s=s_c$. I seek to estimate the contact force $f_{{\langle 34\rangle}}$ that appears in this contact when it closes. By symmetry,  all the results we have derived when the contact $\langle 12\rangle$ was opened and ${\langle 34\rangle}$ was closed also apply to the newly obtained configuration if ${\langle 34\rangle}$ is re-opened and $\langle 12\rangle$ is re-closed.
In particular  the relation $f_{\langle 12\rangle}=\partial (p\delta V(s))/\partial s|_{s=0}$ becomes $f_{\langle 34\rangle}=\partial (p\delta V(s'))/\partial s'|_{s'=0}$, where $s'$ is the distance by which the contact ${\langle 34\rangle}$ is opened. One has $\partial s'/\partial s\equiv -D$  where $D$ is a positive constant of order one by symmetry. Thus $f_{\langle 34\rangle}$ is readily obtained by differentiating Eq.(\ref{3bis}) at $s_c$, and leads to:
\be
\label{8}
f_{\langle 34\rangle}= D\left( \frac{AN\langle f\rangle s_c}{\sigma_0} - f_{\langle 12\rangle}\right)
\ee
Using Eqs.(\ref{4bis}) and (\ref{8}) in the case where $\langle 12\rangle$ is the weakest contact, i.e. $f_{\langle 12\rangle}=f_{min}$, one sees that the violation of inequality (\ref{7}), which implies  $s_c >> s^*_{min}$, lead to the condition $f_{\langle 34\rangle}>> f_{min}$.

I now argue that the creation of a new contact with a force much larger than the typical minimal forces $f_{min}$ will trigger  an avalanche of an extensive size. Closing the contact ${\langle 34\rangle}$ is equivalent to imposing an external dipole of forces - just before they touch-  on the two particles $3$ and $4$ forming this contact, of magnitude ${\vec F}_3=-{\vec F}_4=f_{\langle 34\rangle}{\vec n}_{34}$. The response to such a force dipole does not change the mean contact force $\langle f\rangle$, because the pressure is fixed. However property (d) implies that contact forces throughout the system are changed by some random amount, of order $f_{\langle 34\rangle}$. Since $f_{\langle 34\rangle}$ is much larger than the smallest forces in the system, many  contact forces become negative when the contact ${\langle 34\rangle}$ is formed. A negative contact force would correspond, in the representation of Fig(\ref{f1}), to a negative slope at $s=0$, leading to an instability where the contact opens. The opening of these contacts will lead in turn to new contacts forming, themselves generating some significant noise in the values of contact forces, and triggering new  openings of contact. Such a dynamical process will stop only when inequality (\ref{7}) is satisfied. It is likely that something of this sort takes place each time a packing of hard particles is prepared.

{\it Discussion}:
Imposing the stability of  the contact network  leads to  an inequality between the distribution of forces and the pair distribution function in packings, Eq.(\ref{7}).    In glasses with long range interactions such a stability bound exists, it is  saturated both the equilibrium state \cite{thouless} and in non-equilibrated configurations  \cite{moore,horner} in spin glasses, and nearly saturated in the Coulomb glass  \cite{goethe,markus}.  In the case of random close packings, thermal equilibrium is  not achieved, and the exponent $\theta$ and $\gamma$ may depend on the  system preparation. 
Empirically, for isotropic packings obtained via decompression of soft particles it is found that $\gamma\approx1/2$ \cite{silbert0,Hern}, whereas for packings obtained via compression of thermal hard particles $\gamma\approx0.4$ \cite{donev2} (in the later measurement however rattlers, corresponding to a few percent of the particles that are not jammed, were not taken into account).  On the other hand   $P(f)$ has been extensively studied in the granular matter literature, but with little precision at low force. One accurate measurement of $\theta$ was made in anisotropic jammed packings \cite{edan2} (where the present argument should also hold), and yields $\theta=0.2$. $\gamma$ was not measured in that case however, and the saturation of Eq.(\ref{7}) would correspond to $\gamma=0.44$, a value similar to what is observed in isotropic packings.  Thus the existing measurements are consistent with the  non-linear marginal stability of packings, although more accurate measurements are obviously needed  to test this hypothesis.

Furthermore, marginal stability is a natural explanation for the observation that the response  to an applied shear stress displays
jumps of strain $\delta \epsilon$, which follow a distribution $P(\delta \epsilon)\sim \delta\epsilon^{-1.46}$ \cite{rouxDL}. Such power-law behavior indicates that the the contact network is critical.
 Criticality can be obtained by fine-tuning parameters, such as in the mean-field ferromagnet in random field \cite{Sethna} (whose exponent $3/2$ is interestingly close to the present one),
 or via some kind of self-organized criticality. The marginal stability proposed here is consistent with the second scenario. Along this line of thought, when a packing is formed the dynamics consists of large avalanches of contacts rearrangements. When extensive avalanches are not possible anymore, the dynamics stops rapidly, ensuring that the system remains close to critical state where the packing is marginal.

Finally,  I have focused on hard frictionless spherical particles, which can describe accurately  emulsions \cite{brujic}.  Often however these assumptions  do not apply: in granular matter particles are not perfectly spherical, there is friction and particles can be deformed to some extent. These features move the system away from isostaticity: for example elliptic particles are hypostatic and present floppy modes \cite{zorana,oherne}, whereas friction or softness make the system hyperstatic. In both cases, one expect these systems to behave like isostatic ones below a length scale $l^*$ that diverges near isostaticity \cite{Wyart05}, suggesting that the  proposed description of  the network dynamics  applies on this mesoscopic length scale. One challenge for the future is to connect  the present approach to long wavelength phenomena, for example the emergence of avalanches of localized  plastic events in soft glasses or  the apparition of shear bands in granular matter.

Acknowledgments: it is a pleasure to  thank  A. Grosberg, G. During, P. Hohenberg,  E. Lerner, E. Vanden-Eijnden  for comments on the manuscript. This work has been supported by
the Sloan Fellowship, NSF DMR-1105387, and Petroleum Research Fund \#52031-DNI9. This work was also supported partially by the MRSEC Program of the National Science Foundation under Award Number DMR-0820341.

\bibliography{reference6.bib}{}

\begin{thebibliography}{38}
\bibitem [{foo()}]{footnote1}%
  \BibitemOpen
  \href@noop {} {}\bibinfo {note} {In an isostatic system if $\delta{\vec R}_i$
  is the displacement of some particle $i$ following the opening of a contact
  $\langle 12\rangle$ by some distance $s_{\langle 12\rangle}$, then the change
  of contact force $\delta f_{\langle 12\rangle}$ following the application of
  an external force ${\vec F}_i$ on particle $i$ reads $\delta f_{\langle
  12\rangle}={\vec F}_i\cdot \delta{\vec R}_i/s_{\langle 12\rangle}$.
  Properties (a) imply that $ ||\delta {\vec R}_i||\sim s_{\langle 12\rangle}$,
  thus $\delta f_{\langle 12\rangle}\sim ||{\vec F}_i||$. The same result is
  valid for a force dipole using property (c), which implies that ${\vec F}
  \cdot (\delta{\vec R}_i -\delta{\vec R}_j)$ and ${\vec F} \cdot \delta{\vec
  R}_i $ are of the same order of magnitude.}\BibitemShut {Stop}%
\end{thebibliography}
%

\end{document}